\documentstyle[12pt,aaspp,epsf]{article}
\begin{document}
\def\halfsp{\baselineskip 10truept plus .1 truept}
\def\singlesp{\baselineskip 13truept plus .25 truept}
\def\onehalfsp{\baselineskip 19.5truept plus .3 truept}
\def\doublesp{\baselineskip 26truept plus .4 truept}
\singlesp
\setcounter{tocdepth}{4}
\tableofcontents
\newpage

\title
{18. CLASSICAL SPECTROSCOPY}

\vskip 0.2cm

\author
{J. Bland-Hawthorn\footnote{\it Current Address: \rm Anglo-Australian Observatory, P.O. Box 296, Epping, NSW 2121.}}

\affil
{Department of Space Physics \& Astronomy, Rice University, Houston, TX 77251}

\vskip 5pt
\centerline{and}

\author
{G. Cecil}

\affil
{Department of Physics \& Astronomy, University of North Carolina, Chapel Hill, NC 27599}

\stepcounter{section}
\stepcounter{section}
\stepcounter{section}
\stepcounter{section}
\stepcounter{section}
\stepcounter{section}
\stepcounter{section}
\stepcounter{section}
\stepcounter{section}
\stepcounter{section}
\stepcounter{section}
\stepcounter{section}
\stepcounter{section}
\stepcounter{section}
\stepcounter{section}
\stepcounter{section}
\stepcounter{section}
\stepcounter{section}

\newcommand{\AIRY}{{\mbox{${\cal A}$}}}
\newcommand{\RES}{{\mbox{${\cal R}$}}}
\newcommand{\THRU}{{\mbox{${\cal T}$}}}
\newcommand{\hh}[2]{\noindent{\sl #1 #2}\addcontentsline{toc}{subsection}{#1 #2}}

\subsection
{Introduction.}

This chapter is a basic review of the fundamental principles of modern spectroscopy,
designed to provide a research student access to a widespread field. We begin by
describing the basic properties of spectrometers and emphasize the relative
merits of certain techniques. Attention is given to recent developments in monochromators,
in particular, narrowband and tunable filters. Extensive historical reviews are to be found
elsewhere [1,2]. While spectroscopic techniques continue to evolve, they rely on either
dispersion (refractive prisms) or multi-beam interference (diffractive gratings, interferometers).
Prisms are of great historical importance [3,4,5], but their performance is now surpassed by
transmissive and reflective gratings [6,7,8]. 
At the close of the 19th century, three important
developments occurred. The curved grating removed the need for auxiliary optics, thereby extending
spectroscopy into the far ultraviolet and infrared wavelengths [7]. This was soon followed by the
development of the Fourier Transform [9] and Fabry-Perot [10]
interferometers that have many uses today. Good general discussions can be found in [11,12,13]. 
More specialized treatments are as follows: prisms [11,14,15], gratings [2,16], Fabry-Perot
[17,18] and Fourier Transform interferometers [19,20].

\subsection
{Basic Principles.}

The most useful figure of merit of a spectrometer
is the product of the resolving power ($\RES$) and the throughput ($\THRU$).
The {\it throughput} is defined as ${\THRU} = A\cdot\Omega$ where $A$ is the normal area of the beam
and $\Omega$ is the solid angle subtended by the source. The {\it resolving power} is defined as
$\RES = \lambda/\delta\lambda$ where $\delta\lambda$ is known as the {\it spectral purity}, or
the smallest measurable wavelength difference at a given wavelength $\lambda$. In a properly
matched optical system, the throughput, or equivalently, the flux through the spectrometer depends
ultimately on the entrance aperture and the area of the dispersive element. The theoretical limit
to the resolving power is set by the characteristic dimension of the spectrometer (e.g. prism
base). In practice, it is often advantageous to accept a lower value of $\RES$, for example, by
widening the entrance slit to allow more light to enter the
spectrometer. Indeed, if an observed spectral line is not diffraction limited, the
flux through the spectrometer is inversely related to the resolving power of the system.
Thus, it makes sense to compare the relative merits of spectrometers at the same effective
resolving power. If we match the area of the dispersing element for each technique $-$ prism,
grating, Fabry-Perot etalon, Fourier Transform beam-splitter $-$ we find that the
solid acceptance angles of the latter techniques have a major throughput (Jacquinot) advantage 
over the others.

\hh{    18.2.1}{Throughput Advantage.}
Jacquinot [21] demonstrated the relative merit
of prisms, gratings and Fabry-Perot etalons on the basis of throughput.
The simplest spectrometer comprises a {\it collimator} optic to equalize
the optical path lengths of each ray between the entrance aperture and the {\it disperser}, 
and a {\it camera} optic to reverse the action of
the collimator by imaging the dispersed light onto the {\it detector}.
If the solid angle subtended by the object, or {\it source}, at the distance of the collimator is
$d\Omega$, and the projected area of the collimator is $dA$, the radiant flux
falling on the collimator from a source of brightness $B$ is given by $dF = B\cdot dA\cdot d\Omega$.
The quantity $dA\cdot d\Omega = dF/B$ is simply the throughput or the {\it \'etendue} of the system.
In a properly matched (lossless) optical system, the \'etendue
is a constant everywhere along the optical path, in which case the brightness of the source is equal
to the brightness of the image.
This is easily seen for a simple lens with focal length $f$. A small element $dA$ at
the focal (or object) distance subtending a solid angle $d\Omega$ is brought to a focus
at the image plane distance $f^{\prime}$ with area $dA^{\prime}$
subtending an angle $d\Omega^{\prime}$. The linear magnification,
both horizontally and vertically, is $f^{\prime}/f$.  Therefore
$dA^{\prime}/dA = (f^{\prime}/f)^2$. However, this is compensated exactly by the
change in $d\Omega^{\prime}/d\Omega$ such that $dA\cdot d\Omega = dA^{\prime}\cdot d\Omega^{\prime}$.
Note that the f/ratio of the optic sets the solid angle of the system.
Therefore, it is important to match all optical elements in series such that the
output of one element fills the aperture of the next.
On the basis of throughput alone, the Fabry-Perot is superior to
grating instruments, which in turn have superiority over prisms. However, in
practice, Fabry-Perots are normally used for high resolution observations over
a narrow wavelength interval. A better comparison is with the Fourier Transform
spectrometer (FTS) which, for the same bandwidth, has a much higher throughput than
all slit-aperture spectrometers in the same configuration.

\hh{    18.2.2.}{Resolving Power. }
The resolution of a spectrometer is set by the bandwidth limit imposed by the
dispersing element.  An elegant demonstration using Fourier optics is quoted by Gray [16].
When a beam of light passes through an aperture of diameter $L$, in the far field approximation,
a Fraunhofer diffraction pattern arises.\footnote{\halfsp The radial diffraction pattern is a
{\sf sinc}$^2$
function where ${\sf sinc}\ y = \sin y/y$. Recall that the ${\sf sinc}$ function is the
Fourier Transform of the rectangle (`top hat') function.}  The width of the central
intensity spike is proportional to $\lambda/L$ which is roughly $\delta\lambda$ for most
resolution (e.g. Rayleigh, Sparrow) criteria. Because the dispersing element defines
a finite baseline or aperture, the ultimate instrumental resolution is set by the
diffraction limit.  This has the simple consequence that the highest spectroscopic
resolutions have generally been obtained with large spectrometers.

The theoretical value of $\RES$ is rarely achieved in practice, not least because of
optical and mechanical defects within the instrument. 
The width of the {\it instrumental profile}, i.e., the response of the spectrometer to a 
monochromatic input, must be matched carefully to the size of each detector element 
(or {\it pixel}) and the sampling interval defined by the angular dispersion.
Gray [16] shows that an observed spectrum arises from the product of three functions, viz.,
$S_{\lambda}=B_\lambda\cdot{\rm III}_\lambda\cdot O_\lambda$ where $O_\lambda$
is the original spectrum, $B_\lambda$ is a rectangle function that defines the baseline of the
dispersing element, and ${\rm III}_\lambda$ is a Shah function that describes the regular sampling.
Because the Fourier Transform of the observed spectrum is the convolution of the transform of
the individual functions, discrete sampling causes the transform of the original
spectrum to replicate with a periodicity $1/\delta x$ where $\delta x$ is the sampling
interval. If the original spectrum is undersampled by the spectrometer, the adjacent orders
of the transformed spectra overlap and cannot be disentangled uniquely. This problem of 
{\it aliasing} can be avoided by {\it Nyquist sampling}, i.e., sampling the source spectrum 
at twice the frequency of the highest Fourier component that you wish to study.

\hh{    18.2.3.}{Detector Constraints. } Most contemporary spectrometers employ panoramic, electronic
detectors because they are highly sensitive (more than 80\% of the incident photons can be detected in many
cases), provide digital output, have a linear response over a dynamic range of $10^5$,
and record photons across a two-dimensional field.  Several corporations now routinely manufacture low-noise,
large area charge-coupled devices (CCDs): in the optical, 2048$^2$ 15$\mu$m and
4096$^2$ 7.5$\mu$m pixel arrays are now available; in the infrared, arrays of
1024$^2$ pixels have been fabricated.  The photo-sensitive area is currently limited
to the 10 cm diameter of a silicon wafer, but suitably designed CCDs can be edge-butted to form
large mosaics. A full description of the limitations and capabilities of modern CCDs can be found
in [22]. At optical wavelengths, the largest detector formats are 35 cm photographic plates. The
Kodak TechPan emulsions have a quantum efficiency close to 5\% and an effective resolution
element of roughly 5 $\mu$m.

To minimize the thermal and collisional excitation of electrons (which contribute a `dark
current'), arrays must be cooled to below 30 K for InSb infrared arrays (which therefore require
inconvenient cryogens) or to below 210 K for modern optical CCDs (obtained using thermoelectric
coolers).  Both require water-free atmospheric chambers to prevent frost, which restricts
access to the focal plane.  Uncorrelated noise sources combine in quadrature, so the
effective signal-to-noise ratio counted at each pixel is
\begin{equation}
{\rm S/N} \approx\frac{s_\nu \Delta_\nu t}{\sqrt{s_\nu\Delta_\nu t+2([b_\nu\omega\Delta_\nu+d]t+\sigma_R^2)}}
\end{equation}
for which $\sigma_R$ is the read noise (generated in the output amplifier, in rms electrons);
$s_\nu\Delta_\nu$, $b_\nu\Delta_\nu$, and $d$ are
the number of electrons per second generated by the object, background, and dark current respectively
in a frequency interval $\Delta_\nu$; $t$ is the exposure time (seconds); and $\omega$ is the solid
angle subtended by each pixel.  The factor of two in the denominator assumes that the corresponding
noise sources are measured on separate exposures, then subtracted from the data frame.
Read noise on modern optical CCDs is a few electrons rms; infrared arrays are at least
ten times noisier but their performance can be improved with non-destructive, multiple
read out.  In practice, the S/N ratio will be smaller than predicted by Eq.\ (1)
because multiplicative (`flat field') and additive (`bias offset') gains must be determined
empirically and applied to each detector element [23]. Except at very low and high flux levels,
the CCD is a linear detector so these calibration steps are often quite successful.

\begin{figure}[t]
\label{Hadamard}
\centering
\hbox{
\epsfxsize=4.8in
\epsfbox[40 570 450 705]{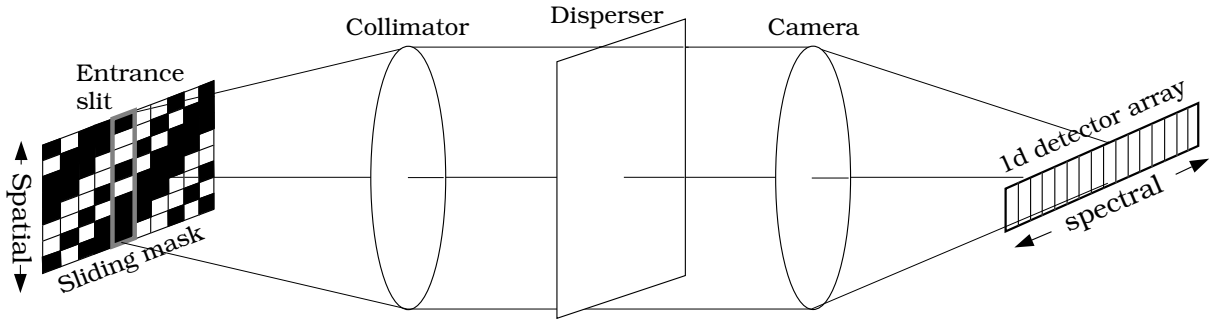}
}
\caption{
\singlesp
A rudimentary Hadamard spectrometer and one-dimensional detector array. The only light to pass
through the vertical entrance slit is through the mask holes aligned with the slit.
After sliding each column over the slit, it is possible to reconstruct the spectrum
at each spatial position along the slit even though the detector does not extend in this
direction.
}
\end{figure}

\hh{    18.2.4.}{Multiplex Advantage. }
In recent decades, the sense of what constitutes a multiplex (Fellgett) advantage has evolved. The
traditional meaning arises from single element or row element detectors which used to prevail at
infrared wavelengths. With a single element detector, a two-dimensional image (either spatial-spatial
or spatial-spectral) was made by scanning at many positions over a regular grid. The same image is
more easily obtained with a one-dimensional detector array after aligning one axis of the image with 
the detector and then shifting the detector in discrete stages along the other axis. For a single
element detector, Fellgett [24] realized that there is an important advantage to be gained by recording 
more than one spectral increment (channel) simultaneously if the signal detection is limited by detector (background) 
noise. If the receiver observes in sequence $n$ spectral channels dispersed by a prism or grating for 
a total exposure time of $\tau$, the S/N ratio within each channel is proportional to $\sqrt{\tau/n}$. 
So, a Fourier Transform device (see \S~18.6), which observes all $n$ spectral channels for the entire 
duration, has a {\it spectral} multiplex advantage of $\sqrt{n}$ compared with conventional slit 
spectrometers. Thus, a multiplex advantage makes more efficient use of the available light.

In principle, all spectroscopic techniques can achieve a multiplex advantage with
the use of a multiplexed or coded {\it aperture mask}. Spectrometers with cylindrical
symmetry (e.g. Fabry-Perots) use circular masks, while slit aperture devices
(e.g. gratings) use rectangular masks.  Fig.\ 1 illustrates the special case of Hadamard
coded apertures. A cyclic mask $H_{ij}$ with $n$ rows is placed at the
entrance slit; a particular column can be aligned with the slit by sliding the mask. In this
way, the dispersing element can be illuminated through each of the mask columns in turn.
A one-dimensional detector array is aligned with the spectral dispersion so that
each spectral channel receives all of the modulated signal through the slit at a discrete
frequency.  For each spectral channel, the modulated signal $M_i = H_{ij} O_j$ can be used
to derive the original signal $O_j$ at each position along the slit.
This assumes that the (square) Hadamard array $H_{ij}$ can be inverted and that the influence 
of systematic errors can be minimized.  The multiplex advantage is roughly $\sqrt{n}/2$ if the 
mask columns have approximately half the holes open.

With the advent of large format, low noise detectors in many wavebands,
the conventional definition of the multiplex advantage is only significant
for the highest spectral resolutions. It has become difficult to
generalize under what conditions a multiplex advantage might prevail [11].
For certain applications, it can be inefficient to match the acceptance solid angle
of the spectrometer to the imaging detector. As an example, optical fibers in astronomy 
--- after major improvements in their blue to near-infrared response --- have afforded a significant
{\it spatial} multiplex advantage in recent years [25,26]. The light from discrete sources over a
sparse field can be collected with individual fibers which are then aligned
along a slit. The multiplex advantage, when compared with conventional slit
spectrometers, is simply the number of fibers if the signal attenuation along the
fiber length is negligible.

\hh{    18.2.5.}{Optics Design. }
All spectrometers make use of the constructive interference of light by dividing the wavefront 
(e.g. gratings), by dividing the amplitude of the wavefront (e.g. Fabry-Perots), or by decomposing
the wavefront into orthogonal polarization components (e.g. Lyots). The design of a spectrometer is 
in essence a competition between a dispersing element that deviates light into different angles
according to wavelength, and optics that focus the light at the detector with minimum aberration.
Two lenses in the optical path are normally sufficient to counteract the dominant aberrations [12].
In most cases, the dispersing element must be illuminated with parallel (collimated) light. 
Hence the acceptance angle (i.e. f/ratio) of the collimating optic must be the same as
that of the primary concentrating optics to transfer all the light. If the collimated beam is too
small, the Jacquinot advantage of the spectrometer is reduced [27,28]. If the collimated beam
is too large, light is lost from the optical system.

It is important to realize that all optical systems lose light at surfaces along the optical path. On
occasion, the scattered light will simply leave the system. More often, the
stray light finds its way back into the optical path to be imaged at the detector
as a spurious `ghost' signal that can be difficult to distinguish from real signals. Scattered
light can also dramatically increase the background signal at the detector, thereby reducing
contrast and setting a limit on the sensitivity that can be reached within a given
exposure time.  The manner in which this happens is specific to the instrument.  In later
sections, we describe a few of the ghost families that occur within gratings and Fabry-Perot
spectrometers.  Anti-reflective coatings (including newly developed coatings whose
index of refraction increases smoothly through their thickness),
aperture stops, baffles and ingenious optical
designs are part of the arsenal to combat these anomalies.

In later sections, we describe diffraction gratings ($\S$18.4), interference filters ($\S$18.5.1),
Fabry-Perot ($\S$18.5.2) and Fourier Transform spectrometers ($\S$18.6). We also discuss recent
technological developments for selecting a bandpass whose central wavelength can be tuned over a 
wide spectral range. The most promising of these developments are those which utilise anisotropic
media, particularly the Lyot filter ($\S$18.5.3) and the acousto-optic filter ($\S$18.5.4).

\newpage
\subsection {Prisms.}
Refractive prisms are no longer in common use as primary dispersers in slit spectrometers, 
although they are frequently used as cross dispersers in high-order spectroscopy and they
play an important role in immersion gratings. However, prisms
highlight some of the basic principles discussed in the previous section. Their
operation is fully specified by Snell's law of refraction and the dispersive properties of the
medium. A light ray incident on the face of a prism with refractive index $n_\lambda$ deviates by an angle
$\delta = i-r$, where $i$ is the angle to the face normal, and $r$ is the refracted angle. If
we take the external medium to be air, then $\sin i = n_\lambda\ \sin r$.
For a restricted wavelength range, the Hartmann dispersion formula,
$n_\lambda = a + b (\lambda-c)^{-1}$, provides a good approximation to many materials in the Schott
glass catalog. The constants $a$, $b$ and $c$ depend on the material.

\begin{figure}[t]
\label{PrismGrating}
\centering
\hbox{
\epsfxsize=3.95in
\epsfbox[40 130 450 630]{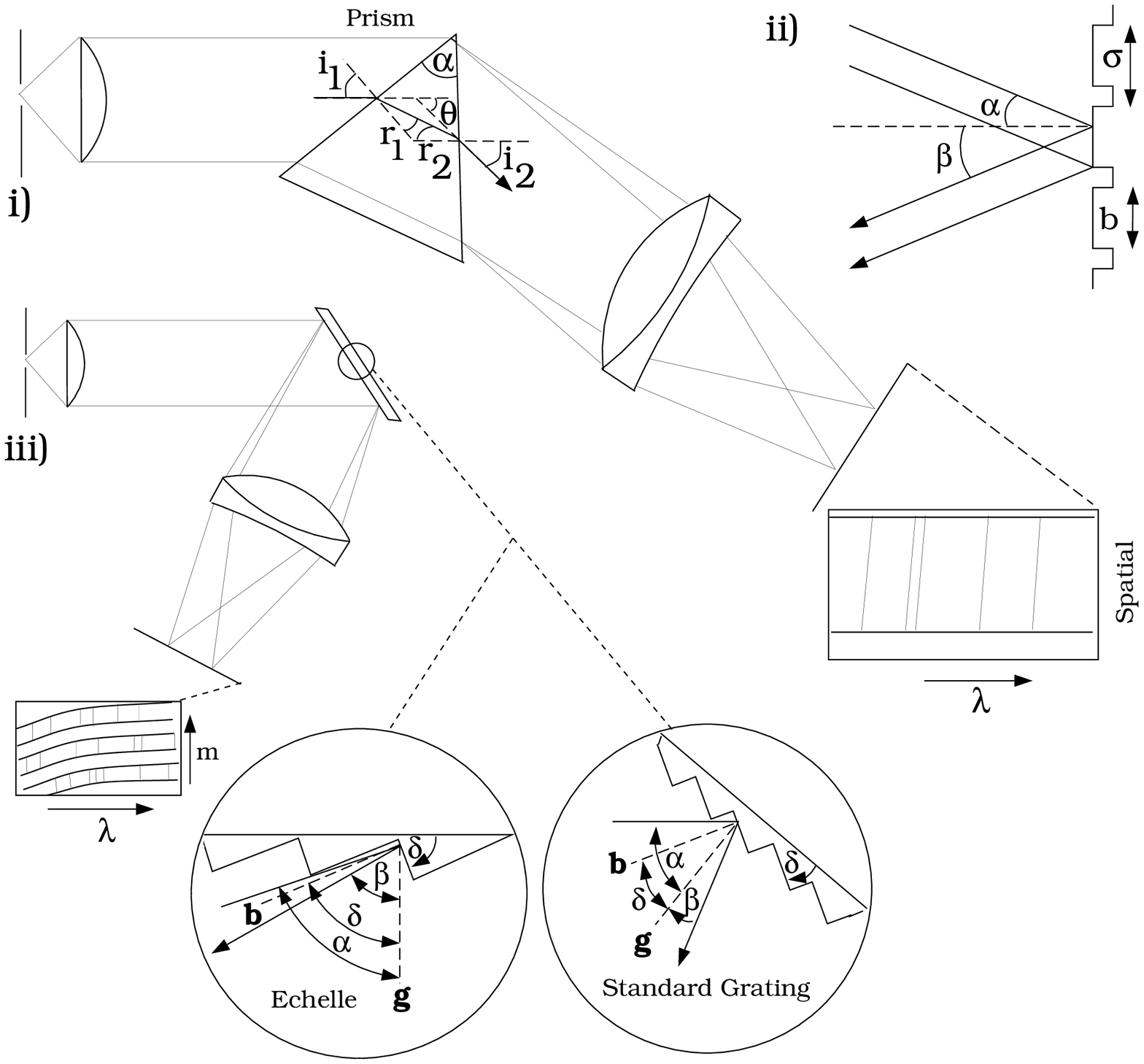}
}
\caption{
\singlesp
(i) A rudimentary prism spectrometer.
(ii) Angle and groove definitions for a conventional, reflective diffraction grating.
(iii) A blazed reflection-grating spectrometer. The
standard grating and echelle arrangements are shown as insets, where the dashed-line
vectors {\bf g} and {\bf b} are the normals to the grating and blaze facets, respectively.
The spectral format of the echelle on the detector assumes that the different orders have
been displaced as shown by auxiliary optics.
}
\end{figure}

Fig.\ 2(i) illustrates the angle convention for a ray passing through a prism. We note that
$\delta_1 = i_1-r_1$, $\delta_2 = i_2-r_2$, and the prism apex angle, $\alpha = r_1+r_2$, such that
the deviation is given by $\theta = \delta_1+\delta_2 = i_1+i_2-\alpha$ or equivalently
\begin{equation}
\theta(\alpha,\lambda,i_1) = i_1 - \alpha + \sin^{-1}\left( n_{\lambda} \sin[\alpha-\sin^{-1}(n_{\lambda}^{-1}\sin i_1) ] \right)
\end{equation}
We can approximate the angular dispersion $d\theta/d\lambda$ by finding the gradient between two
discrete wavelengths $\lambda_0$ and $\lambda_1$.  If we plot
$(\theta(\alpha,\lambda_1,i_1)-\theta(\alpha,\lambda_0,i_1))/(\lambda_1-\lambda_0))$ over a range of
$i_1$ for a flint prism, say, we find that it increases with $\alpha$ to a theoretical maximum at which
point $\alpha \approx 74^{\circ}$ and $i_1 = 90^{\circ}$ [14]. This corresponds to a ray at glancing
incidence on the first face followed by a symmetric passage or, equivalently, minimum deviation
through the prism. Alternatively, we can substitute $i_1$ with $r_1$ and differentiate to find
\begin{equation}
{{\partial}\over{\partial r_1}}\theta(\alpha,\lambda,r_1) = {{n \cos r_1}\over{\sqrt{1-n^2 \sin^2 r_1}}} - {{n \cos(\alpha-r_1)}\over{\sqrt{1-n\sin^2(\alpha-r_1)}}}
\end{equation}
Thus, minimum deviation occurs when $r_1 = r_2 = {{1}\over{2}}\alpha$ which corresponds to a symmetric
passage through the prism.  At minimum deviation, we let $i = i_1 = i_2$ and $r = r_1 = r_2$. After
substituting $\theta = 2i-r$ and $\alpha = 2r$, one finds
\begin{equation}
{{d\theta}\over{d\lambda}} = {{2\sin\frac{1}{2}\alpha}\over{\sqrt{1-n^2\sin^2\frac{1}{2}\alpha}}} {{dn}\over{d\lambda}}
\end{equation}
The maximum apex angle is not used in practice because
most of the light is reflected at the first
surface. Manufacturers cut prisms to apex angles (e.g., $30^{\circ}, 45^{\circ}, 60^{\circ}$)
that minimize wasted glass. The $60^{\circ}$ prism is a good compromise for which the angular dispersion
is approximately $n\ dn/d\lambda$. This function rises rapidly to large and small wavelength
cut-offs [11,21]. Thus, prisms disperse most efficiently at their absorption limits and therefore find use
from ultraviolet to infrared wavelengths (100 nm$-60 \mu$m). A major disadvantage of prisms
is the strongly non-linear angular dispersion.  If the collimated beam is matched to the prism, it follows
from Eq.\ (4) that $\RES = B\ d\theta/d\lambda = L\ dn/d\lambda$ where $L$ is the length of the prism base
and $B$ is the beam diameter.  At grazing incidence, $B=0$ and therefore $\RES=0$.
Finally, for a large 10 cm glass prism, $dn/d\lambda \approx 10^{-4}$ nm$^{-1}$,
and $\RES\approx 10^4$ which is roughly the practical limit of prism spectrometers.

\subsection
{Diffraction Gratings}
If a plane wave of wavelength $\lambda$ is incident at an angle $\alpha$ to the perpendicular
of a periodic grating with groove spacing $\sigma$, the outbound reflected beams at an angle
$\beta$ interfere constructively according to the {\it grating equation}
\begin{equation}
\sigma(\sin\alpha+\sin\beta) = m\lambda
\end{equation}
where $m$ is the order of interference and the angle conventions are illustrated in Fig. 2(ii). It
is common practice to make $\alpha > \beta$ to minimize scattered light. The angular dispersion follows
by differentiating this equation
\begin{equation}
\frac{d\beta}{d\lambda} = \frac{m}{\sigma\cos\beta}
\end{equation}
In many applications $\beta>30\deg$, which causes the angular dispersion to be slightly non-linear
and therefore the wavelength scale at the detector plane must be calibrated from a reference spectrum.
Standard ruled gratings have $\sigma$ = 1/600 to 1/1200 mm and are normally used in relatively low
order ($<5$). The longest wavelength accessible with a grating is $2\sigma$,
so gratings in the infrared
tend to be coarsely ruled.  For a grating length $L$, the maximum resolving power increases as
${\cal R} = Lm/\sigma$ and can exceed $10^5$. The spectral purity, $\delta\lambda$, depends on the
collimator focal length, $f_{\rm coll}$, such that
\begin{equation}
\delta\lambda = -\cos\alpha\frac{w}{f_{\rm coll}}\frac{\sigma}{m}
\end{equation}
where $w$ is the width of a detector element.

Fourier optics demonstrates that the flux distribution in the focal plane is the Fourier
power spectrum of the transmission (or reflection) function over the collimated beam.
The grating acts as a filter that blocks all spatial frequencies except those associated with its groove
frequency and the spatial frequency content of each groove [16].
The number of grooves and the maximum pathlength difference is set by the grating length $L$.
The rectangle function that defines each grating groove produces a broad diffraction envelope 
(called the {\it blaze function}) that modulates the
flux of each spectral order, and whose width increases as the groove narrows (cf. Fig.\ 3). If we ignore the
entrance slit, the wide rectangle function that defines the grating produces a high-frequency diffraction 
envelope (or {\it cluster}) at each spectral order which is modulated by the broad diffraction envelope. 
The width of each
cluster decreases as the overall length of the grating increases.  The maximum intensity of the secondary peak
in a cluster is a fraction $N^{-2}$ of the central peak, where $N$ ($=L/\sigma$) is the total number of
grooves in the grating.  Note that the discontinuities in optical path length produced by the
light-blocking strips between the slits or the steps between the mirrored facets 
provide the high spatial
frequencies that are essential for shaping the profile by interference.

\noindent
\hh{    18.4.1.}{Grating Fabrication. }
Plane reflective gratings are constructed by ruling with a diamond tool
either an aluminum or a gold coating on a low-expansion glass substrate.
To reduce costs, many epoxy-resin replicas are made from each glass master.  Wear on the tool
and changes in the operating environment during the grooving process limit the maximum ruled
area to about $35\times45$ cm, but gratings can be mosaiced in much the same way as silicon
chips (\S~18.2.3).  It takes several weeks of cutting at 10 grooves min$^{-1}$
to fabricate a grating, during which no distance should drift by more than 20 nm.
The actual resolution attained with a grating is smaller than calculated due to slow
variations in the ruling density that can degrade coherence, broaden the wings of the
instrumental profile, and produce spurious spectral line and continuum
features at the detector [29, 30]. Periodic deviations in the engraved facets from flaws
in the ruling engine produce periodic spectral features near bright
spectral lines (Rowland ghosts). When one or more periodic deviations are present, interference
of the ghost diffraction patterns can generate spurious lines far from the original
spectral feature (Lyman ghosts). 

Holographic gratings do not suffer from these mechanical constraints and therefore
can be fabricated with much higher groove densities (3500$-$6000 grooves mm$^{-1}$).
They are formed by imaging the interference pattern of a
laser-fed Fizeau, Michelson (see $\S$18.6) or Twyman Green interferometer
onto a glass plate that has been covered with
photo-resistive emulsion.  The unexposed areas are etched away in an acid bath,
and the resulting sinusoidal surface undulations form a grating.
The profile of each facet must be squared to maintain high efficiency [31].
Another advantage of these gratings is that the
astigmatism of an off-axis spectrograph can be compensated for by shaping the wavefronts of
the interfering beams. However, all gratings produce at least two kinds of
stray light [32]: a Lorentzian component predicted by diffraction theory, and a 
Rayleigh-scattered component due to microscopic surface defects.

\noindent
\hh{    18.4.2.}{Influence of the Entrance Slit. }
The entrance aperture serves a dual function in slit spectrometers.  First, it restricts flux to a
particular region of the source and ensures that the flux is dispersed onto a uniform and
low-level background that is imaged
by the detector.  Second, the recorded spectral lines are inverted images of the
slit width, convolved by the grating profile and further broadened by aberrations within the optics.
The choice of a slit width is a compromise between reduction of the source
intensity or the intrinsic resolution of the grating.
To avoid light loss by diffraction, the width cannot be reduced below about five times
the operating wavelength. In practice, it is hard to make an adjustable slit narrower than
10 \micron\ that maintains parallelism.

The slit may actually be an aperture mask with many separate openings (slitlets) distributed across
the focal plane but arranged so that spectra do not overlap at the detector.  Even more flexibility is
possible by positioning many optical fibers at widely distributed points in the focal plane [25,26].  Such
`fiber feeds' have been attached to existing spectrometers to greatly increase their efficiency for certain
projects.  One end of the fibers can either be attached to precut holes in a custom
mask or can be moved to arbitrary
positions by a robot arm.  The other ends can then be lined up along the spectrograph slit so that
the spectra do not overlap whatever the fiber position. It is common practice to reserve some of the
fibers for direct observations of the background signal.  If the source does not occupy the entire
length of a slitlet, aperture masks produce better background sampling.  This is because each fiber has a
slightly different response: a fiber that sees only the background does not exactly match an adjacent fiber that
is illuminated by the combined source and background signal.  Fibers are also used as `image scramblers' when
it is important to minimize spectral artifacts that are introduced as a point source moves around in a large
slit.

The collimator f/ratio is fixed to that of the primary collecting optics for full illumination, while its
diameter must be that of the grating to maintain the grating resolution. The ratio of the camera and the
collimator focal lengths, $f_{\rm cam}/f_{\rm coll}$, demagnifies the scale along the slit. The camera focal
length is set to reduce the slit image width $W$ down to the width of two detector elements $2w$ to satisfy the
sampling theorem (\S~18.2.2), where
\begin{equation}
w = -\frac{\cos\alpha}{\cos\beta}\frac{f_{\rm cam}}{f_{\rm coll}} W
\end{equation}
The {\it anamorphic factor} $\cos\alpha/\cos\beta$ arises from the different beam sizes seen by the camera and
the collimator. It is common for this factor to be less than unity to maximize spectral resolution, in which 
case
the grating is more face-on to the camera than to the collimator.  The camera diameter is also matched to the
grating length which now fixes the f/ratio. Speeds faster than f/2 usually require catadioptric (Schmidt) 
cameras which have fairly inaccessible focal planes, although dioptric cameras are preferred if the 
detector obscures a significant fraction of the beam. Solid glass Schmidt cameras are often used for fast 
systems because the higher index of refraction allows the f/ratio to increase by the same factor.

In many applications, the input scale is imposed by the source. This is particularly
the case for large-aperture ($>3$ m diameter) astronomical telescopes. The main optics 
produce an image scale at the
entrance slit of 206265/($\rho D_{\rm pr}$) (which is typically $\ \la\ $10 angular seconds of 
arc mm$^{-1}$), where $D_{\rm pr}$ (in mm) and $\rho$ are the
diameter of the primary optic and its f/number, respectively.  
At even the best sites, images are blurred by
turbulence at 8-10 km altitude to angular diameters of $\ga$0.6 seconds of arc,
which forces both slits and fibers to be $\ga$100 \micron\ wide
in order to pass most of the light.  Thus, for high values of \RES, a large, high density grating is required
to compensate for the wide slit.  This in turn requires large-diameter optics and a long focal length for the
collimator.  In addition, because CCD pixels are rarely larger than 25 \micron, the camera f/ratio must be
about four times faster than the collimator to demagnify the image.  Cassegrain beams commonly have $\rho=7$
which implies very fast, complex, and expensive camera optics.  New generation 8-10 m telescopes will use
low-order adaptive optics to partially correct the distorted stellar wavefront for atmospheric blurring. 
This will reduce the core of the stellar image to $\approx0.2$ seconds of arc
in diameter, allowing a smaller slit, grating, and optics with minimal light loss. 
A recent development has been to segment the telescope mirror (pupil image) or the focal plane image
with microlens arrays. These generate many sub-images which are brought to focus onto a fiber bundle 
prior to the slit allowing a much simpler camera design.

\hh{    18.4.3.}{Transmission Grating. }
Transmission gratings are used mostly in conjunction with a prism (`grism') or a lens (`grens'),
either in contact or air-spaced. Transmission gratings are commonly used as slitless systems 
which are only practical for spectroscopy of point sources on a weak background signal.
Because no auxiliary optics are required,
they can often be incorporated into an existing optical train to
provide some wavelength selection while maintaining high throughput.
The aberrations associated with a non-collimated beam (principally coma) can be minimized
with suitable grating or detector tilts if the linear dispersion is comparatively low.
Note that the spectra are dispersed across the field of view. In slitless systems, the spectra are 
superposed on a variable background signal, and will overlap if the field is crowded with sources.

\hh{    18.4.4.}{Blazed Grating. }
Most high-efficiency gratings are reflective and have peak efficiencies near 80\%.
As the groove width $\sigma-b$ is decreased (see Fig.\ 2(ii)), the grating acts more like
a plane mirror and concentrates the light
into the non-dispersed $m=0$ order (Fig.\ 3). Time delays must
be introduced across the grating to shift the peak of the diffraction envelope to an angle
that corresponds to dispersive non-zero orders.  Such {\it blazing} is easy to do on a
reflective grating by grooving at an angle $\delta$ to its normal.
The change in effective width of the grooves broadens the grating diffraction
peak asymmetrically, with a more abrupt decline in efficiency on the smaller wavelength
side. Once again, the blaze function has the form ${\sf sinc}^2 \gamma$ for which
\begin{equation}
\gamma = \frac{\pi\sigma\cos\delta}{\lambda}[\sin(\beta-\delta)+\sin(\alpha-\delta)].
\end{equation}
Grating manufacturers quote the blaze wavelength $\lambda_0$ 
for $m=1$ and $\alpha = \beta = \delta$ (Littrow configuration);
$\lambda_0$ and spectral order $m$ are related by $ \gamma = m\pi(\lambda_0-\lambda)/\lambda$.
However, the required exit angle $\beta_{\rm b} \neq \alpha$, which shifts the blaze
to the (typically 10\%) smaller wavelength
\begin{equation}
\lambda_{\rm b} = \lambda_0 \cos\frac{1}{2}(\alpha-\beta_{\rm b})
\end{equation}
The wavelength of the blaze peak $\gamma = 1$
is $\lambda_{\rm b} = 2 \sigma \sin \delta \cos(\alpha - \delta)/m$
when $\alpha+\beta = 2 \delta$.
The blaze curve drops to 40\% of its peak value at the wavelengths
$\lambda_\pm = m\lambda_0/(m\mp\onehalf)$.

\begin{figure}[t]
\label{Transmission}
\centering
\hbox{
\epsfxsize=3.5in
\epsfbox[0 100 375 350]{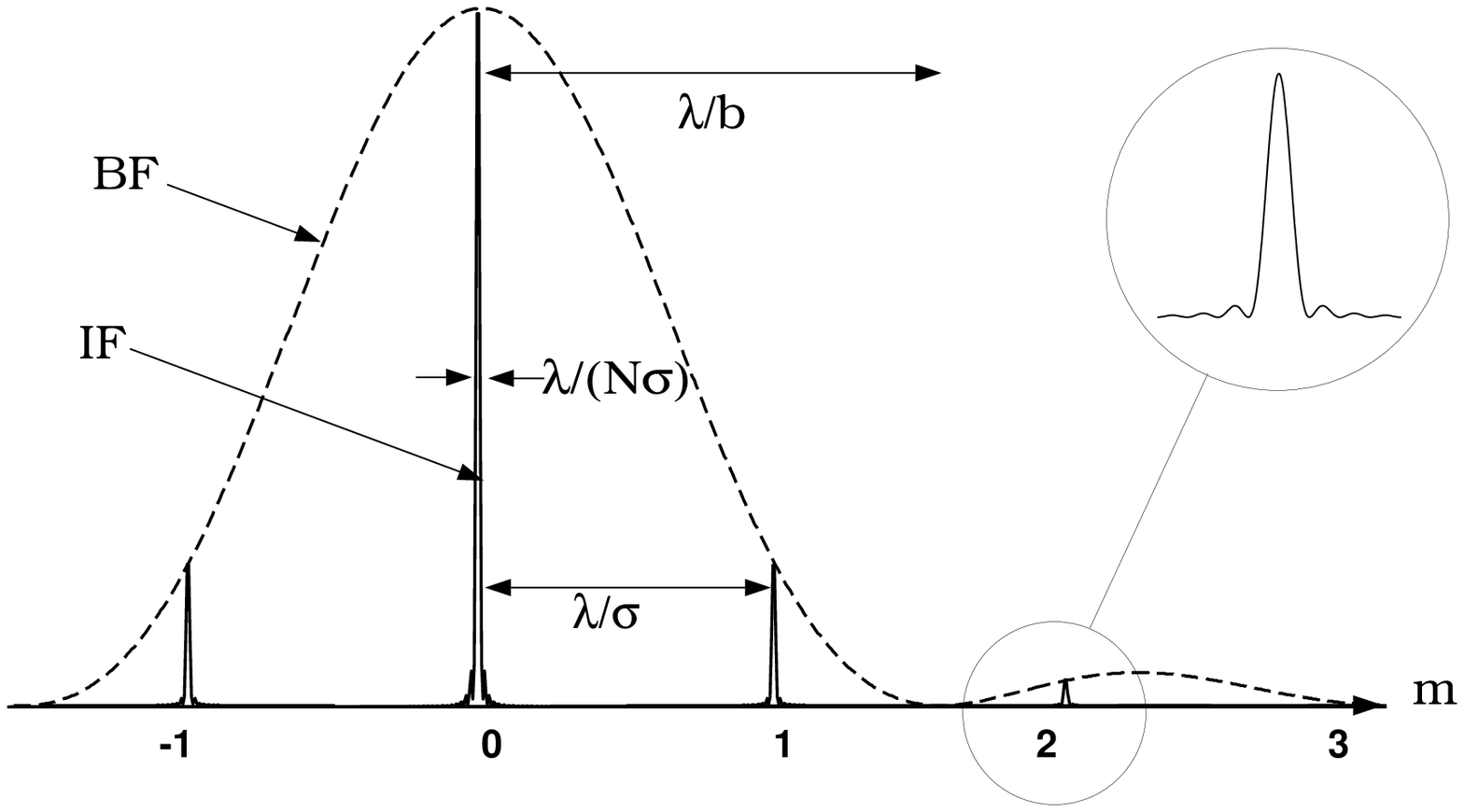}
}
\caption{
\singlesp
Instrumental response of a diffraction grating with blaze angle $\delta = 0$.
BF is the {\sf sinc}$^2$ blaze function and IF is the interference response to a
monochromatic source.  When $\sigma = b$, only the $m=0$ order passes flux because the other
peaks coincide with the BF zeros.
}
\end{figure}

At high order, the wavelength range spanned by the blaze curve is small.
When there is a need for large wavelength coverage with moderate resolution
(${\cal R} \approx 10^{4-5}$), the gratings are operated in the {\it echelle} mode (\S 18.4.5).
Off-the-shelf gratings are available with the blaze peak at one of several strong spectral lines.
Because the blaze function is quite strongly peaked at low orders, the ability to manipulate the
orientation of plane gratings accurately is an important design goal for an efficient spectrometer.
This capability is nontrivial to provide within the sealed cryostat of an infrared spectrometer. 
In the infrared, it is also hard to eliminate unwanted higher orders that can extend into the visual.

Because the grating constitutes a series of stepped mirrors, the blaze efficiency depends
strongly on the polarization state [16] and strong
discontinuities with amplitudes $\ga$20\%\ (called Wood's anomalies) are
often present close to the edges of the blaze distribution $\lambda_\pm$.
Modern astronomical telescopes usually mount spectrometers at the Nasmyth foci, which
are fed by a 45\deg\ tertiary mirror.  Reflection off this mirror induces or alters
the polarization of the incident light, and
the varying polarization angle after the further reflection off the grating
greatly complicates spectrophotometric calibrations.

\hh{    18.4.5.}{Echelle Grating. }
While echelles have a low density of grooves ($\approx$80 mm$^{-1}$), they are operated
at a large spectral order $m\approx200$ and high tilt to maintain a large path difference.
The adjacent orders overlap, so they are cross-dispersed by directing the light through a 
prism or secondary grating that is oriented perpendicular to the echelle.
In this way the different spectral orders are distributed as curved arcs across a two-dimensional
detector array. The main advantage of an echelle is that most of the detector area is
used to record spectra (several orders $m$ simultaneously, see Fig.\ 2(iii)).
The large tilt of an echelle means that a rectangular mosaic of gratings
is necessary for full illumination by a large collimated beam.  All of the separate
glass substrates must have identically low coefficients of thermal expansion (e.g. Schott Zerodur 
or Corning ULE glass) to ensure a high \RES\ at the end of the several-week-long
fabrication cycle of the gratings.

Spatial information along the entrance slit is limited
by the small separation between spectral orders.
During data reduction, the different orders are extracted then overlapped
to yield one spectrum that can span the entire sensitivity range of the detector while
maintaining \RES\ $\approx10^{4-5}$.
The blaze peaks when $m\lambda = 2 \sigma\sin\alpha$, which corresponds to many
wavelengths with adjacent spectral orders because $m$ is large.
The blaze curves of the different orders introduce undulations in the combined
spectrum with large amplitude which must be removed by recording the
known continuous spectrum of a calibration source with the same setup.
The variations in S/N ratio along the spectrum can complicate the data analysis.

\hh{    18.4.6.}{Curved Grating. }
Curved gratings are used to avoid focus degradation from
differential chromatic dispersion across a large wavelength range, when operating in
a wavelength region where lenses are ineffective and mirror reflectances are low, or when
it is desirable to make the spectrometer as compact as possible.
The grating surface is figured to act as a collimator, a camera, or both.
The gratings are one-of-a-kind and are therefore expensive, with limited flexibility in the
choice of camera focal length to alter the projected slit widths and spectral coverage.
In addition, because the grating is an off-axis mirror, it introduces astigmatism
unless an auxiliary mirror is introduced.  The magnitude of the astigmatism increases
with the length of the rulings, and the entrance slit must be aligned precisely with
the ruling direction to avoid degrading the spectrograph focus.
Further discussion of grating mounts and spectrograph aberrations can be found in
[11] and [12].

\hh{    18.4.7.}{Immersion Grating. }
While transmission gratings ($\S$18.4.3) have been largely superseded by other grating types 
(e.g. $\S$18.4.4), prisms have found important uses in the context of immersion gratings [33].
Like the transmission grating, the ruling is placed on the downstream face of the prism, but
now the collimated beam is reflected and refracted along a direction close to the original path.
The principle here is that the resolving power of the grating is increased by the refractive
index of the medium [34]. Immersion gratings are particularly useful at infrared wavelengths
where high index materials are readily available. Anamorphic immersion gratings [35] increase 
the resolving power still further by utilising highly wedged prisms to enhance the anamorphic 
factor of the collimated beam ($\S$18.4.2).

\subsection
{Multiple-beam Interferometers}

\hh{    18.5.1.}{Interference Filter. }
These monochromators allow a narrow spectral band-pass to be isolated. The principle relies on a
dielectric spacer sandwiched between two transmitting layers (single cavity). The transmitting 
layers are commonly fused silica in the ultraviolet, glass or quartz in the optical, and water-free
silica in the infrared. Between the spacer and the glass, surface coatings are deposited by 
evaporation which partly transmit and reflect an incident ray. Each internally reflected ray shares
a fixed phase relationship to all the other internally reflected rays. For a wavelength $\lambda$ 
to be transmitted, it must satisfy the condition for constructive coherence such that, in the 
$m$th order,
\begin{equation}
m\lambda = 2\mu l\cos\theta_R
\end{equation}
where $\theta_R$ is the refracted angle within the optical spacer. The optical gap $\mu l$ is
the product of the thickness $l$ and refractive index $\mu$ of the spacer.
An interference filter is normally manufactured at low order so that neighboring orders spanning
very different wavelength ranges can also be used [36]. Additional cavities, while expensive, can
be added to decrease the band-pass or to make the filter response more 
rectangular in shape.
Either the glass material or an absorptive broadband coating is normally sufficient to block 
neighboring orders.

Filter manufacturers normally provide data sheets that describe operation at room temperature and
in a collimated beam.  If the filter is used in a converging beam, the band-pass broadens 
asymmetrically and the peak transmission shifts to shorter wavelengths. In the collimated
beam, the peak transmission shifts to smaller wavelengths by an amount which depends on
the off-axis angle.  In either beam, the wavelength response of
the interference filter can be shifted slightly (tuned) to shorter wavelengths with a small tilt
of the filter to the optical axis.  If $\lambda_I$ is the wavelength of a light ray incident at an 
angle $\theta_I$ (Fig. 4(i)), then from Snell's Law and equation (11), it follows that
\begin{equation}
\left(\frac{\lambda_I}{\lambda_N}\right)^2 = \left[1-{{\sin^2 \theta_I}\over{\mu^2}}\right]
\end{equation}
for which $\lambda_N$ is the wavelength transmitted at normal incidence.
To shift the band-pass to longer wavelengths,  one must increase the filter temperature
and, typically, one can achieve 0.2\AA\ K$^{-1}$.

A more versatile approach in constructing narrowband filters is to use dielectric, multi-layer
thin film coatings.  A highly readable account of optical interference coatings is provided by 
Baumeister and Pincus [37]. One of the most successful of these is the quarter-wave stack in which
alternate layers of high and low refractive index media are used. Through judicious combinations 
of refractive index and layer thickness, it is possible to select almost any desired bandwidth, 
reflectance and transmittance [38]. However, filters with band-passes narrower than $\lambda$/100 
are difficult and costly to manufacture.

In the next section, we discuss scanning Fabry-Perot etalons which, for the purposes of this
review, use air gap spacers and are routinely operated at low and high order. There exists 
another class of interference filters that is essentially a single cavity Fabry-Perot with a
solid dielectric spacer. These {\it etalon filters} employ a transparent piezo-electric spacer,
e.g. lithium niobate, whose thickness and, to a lesser extent, refractive index can be 
modified by a voltage applied to both faces. Once again, tilt and temperature can be
used to fine tune the band-pass if it is important to keep the piezo-electric voltages modest. 
High quality spacers with thicknesses less than a few hundred microns are difficult to 
manufacture, so that etalon filters are normally operated at high orders of interference.

Any air-glass interface
reflects about 4\% of the incident light. This can be significantly reduced (1\% or better) 
by the application of an anti-reflective coating.  In its simplest form, this constitutes a single 
$\lambda/4$ layer of, say, MgF$_2$ whose refractive index is close to $\sqrt{\mu}$ (Fig. 4(i)). A 
multi-layer `V-coat' with alternating layers of TiO$_2$ and MgF$_2$ dielectric coatings can reach 
\onequarter\% reflectivity at specific wavelengths. Many advances in spectroscopic techniques in 
recent years have arisen from the refinement of multi-layer coatings. The most recent developments 
have used rare earth oxides and very thin (5\AA) metallic layers.  However, coating performance 
is currently limited by the availability of pure transparent dielectrics with high refractive indices. 

\begin{figure}[h]
\label{Filters}
\centering
\hbox{
\epsfxsize=4.8in
\epsfbox[-50 390 450 740]{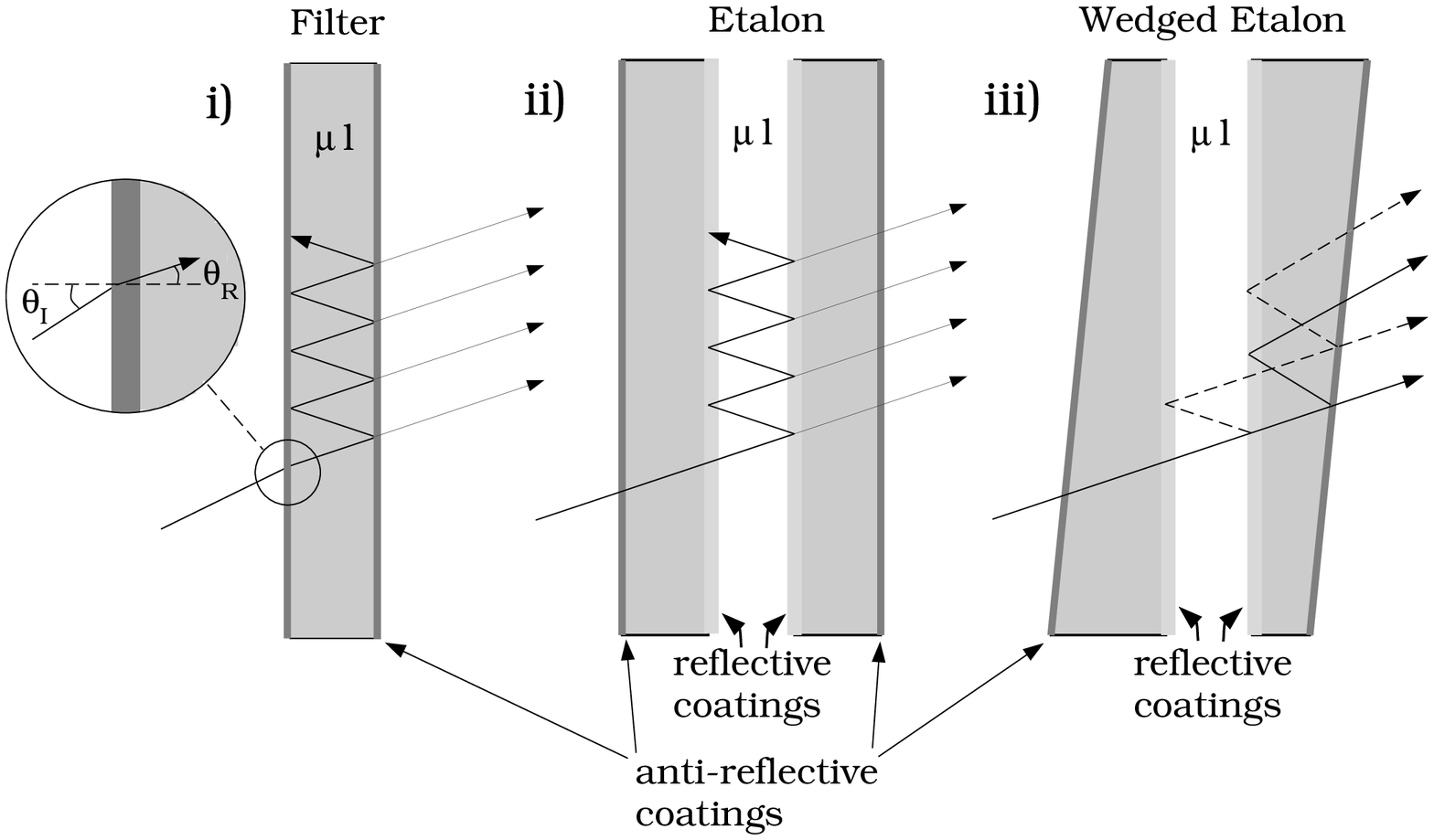}
}
\caption{
\singlesp
(i) Interference filter: the internal structure is not shown. (ii) Fabry-Perot etalon. (iii) 
Wedged Fabry-Perot etalon.
}
\end{figure}

\hh{    18.5.2.}{Fabry-Perot Spectrometer. }
There exists a wide class of multiple-reflection interferometers [36]. With the
exception of interference filters, the Fabry-Perot remains the most popular
in this class. Fabry-Perots are not true monochromators in the sense of interference
or acousto-optic filters (\S18.5.3) and, indeed, require an auxiliary low resolution 
monochromator
to block out neighboring orders. Fig.\ 5 shows the simple construction of a Fabry-Perot
spectrometer: an etalon $d$ is placed in a collimated beam between a collimator $c$ and a
camera lens $e$. The internal structure of the etalon is shown in Fig.\ 4.
The image plane detector $g$ resides within a cooled detector housing; light
passes through a window $f$.  An interference filter $b$ is normally placed close
to the focal plane $a$ or in the collimated beam (Fig. 5(ii)).
The etalon comprises two plates of glass kept parallel over a small separation $l$ (Fig. 4(ii)),
where the inner surfaces are mirrors coated with reflectivity $\Re$. The transmission of the
etalon to a monochromatic source $\lambda$ is given by the Airy function
\begin{equation}
\AIRY = \left( 1+{{4\Re}\over{(1-\Re)^2}}\sin^2 (2\pi\mu l\cos\theta/\lambda) \right)^{-1}
\end{equation}
where $\theta$ is the off-axis angle of the incoming ray and $\mu l$ is the optical gap.
The peaks in transmission occur at $m\lambda = 2\mu l\cos\theta$
where $m$ is the order of constructive coherence. From this equation, it is clear that
$\lambda$ can be scanned physically in a given order by changing $\theta$
(tilt scanning), $\mu$ (pressure scanning), or $l$ (gap scanning). Both tilt and pressure
scanning suffer from serious drawbacks which limit their dynamic range [39]. With the
advent of servo-controlled, capacitance micrometry [40,41], the performance of gap scanning
etalons surpasses other techniques. These employ piezo-electric transducers
that undergo dimensional changes in an applied electric field, or develop an electric
field when strained mechanically.  In practice, the etalon plates are built to a characteristic
separation or {\it zeropoint gap} about which they move through small physical
displacements. The scan range is limited by the dynamic range of the transducers to roughly
$\pm$5 $\mu$m.  For an arbitrary etalon spacing in the range of a few
microns to 30 mm, it is now possible to maintain plate parallelism to an accuracy of
$\lambda/200$ while continuously scanning over several adjacent orders
about the zeropoint gap [42].

The resolution of the Fabry-Perot is set primarily by the reflective coating that is applied
to the etalon plates. The {\t reflective finesse} is defined as $N_R = \pi\sqrt{\Re}/(1-\Re)$.
Ideally, this constant defines the number of band-pass widths (spectral elements) over which an
etalon may be tuned without overlapping of orders.  In other words, the {\it finesse} is the ratio of
the inter-order spacing ({\it free spectral range}) and the spectral purity ($\Delta\lambda/\delta\lambda$).
The etalon is fully specified once the free spectral range and spectral purity are decided [43,44]
because this sets the order of interference ($m=\lambda/\Delta\lambda$), the resolving power
($\RES = \lambda/\delta\lambda = m N_E$), and the zeropoint gap ($l_0 = \lambda^2/(2\mu \Delta\lambda$)).

However, what is measured by the instrument is the {\it effective finesse} $N_E$ which is always less 
than or roughly equal to the theoretical value of $N_R$. There are two effects that serve to
degrade the theoretical resolution. If the beam passing through the etalon is not fully collimated,
the instrumental function broadens and its response shifts to smaller wavelengths. This is analogous
to what happens to an interference filter in a converging beam [45]. The induced profile degradation is
measured by the {\it aperture finesse}, or $N_A = 2\pi/(m\Omega)$ where $\Omega$ is the solid angle set by the
f/ratio of the incoming rays. The profile degradation is negligible in beams slower than f/15.
Another important source of degradation arises from defects in the flatness of the etalon plates.
The {\it defect finesse} is given by $N_D = \lambda/(2\ \delta l)$ where $\delta l$ is the rms
amplitude of the micro-defects. In practice, the defect finesse will also include
terms for large-scale bowing and drifts in parallelism [39]. In essence, to realize an
effective finesse of 50 requires that the spacing and parallelism be maintained to $\lambda/100$. 
In summary, the effective finesse is set by the characteristics of the
etalon surfaces, for which we can write $1/N_E^2 = 1/N_R^2 + 1/N_A^2 + 1/N_D^2$.

There are two primary methods for extracting spectral information from imaging Fabry-Perot
spectrometers. The {\it interferogram} or {\it areal method} measures the perturbations in the
radii of the interference rings to determine spectral differences across the field of 
observation. Roesler
[46] shows that if the radius $r$ of the $k$th ring has been perturbed by $\delta$,
then the corresponding fraction of a free spectral range is
$(r_{k+\delta}^2-r_k^2)/(r_{k+1}^2-r_k^2)$. While this method is of historical importance,
it is now more convenient to assemble the image frames into a three-dimensional data stack and to form
spectra along one axis. In this {\it spectral method} the etalon is scanned to
obtain a sequence of narrowband images taken over a fixed grid of etalon spacings. As the
gap is scanned, each pixel of the detector maps the convolution of the Airy function and the filtered
spectrum at that point [43,44].

\begin{figure}[t]
\label{Ghosts}
\centering
\hbox{
\epsfxsize=4.5in
\epsfbox[60 330 450 650]{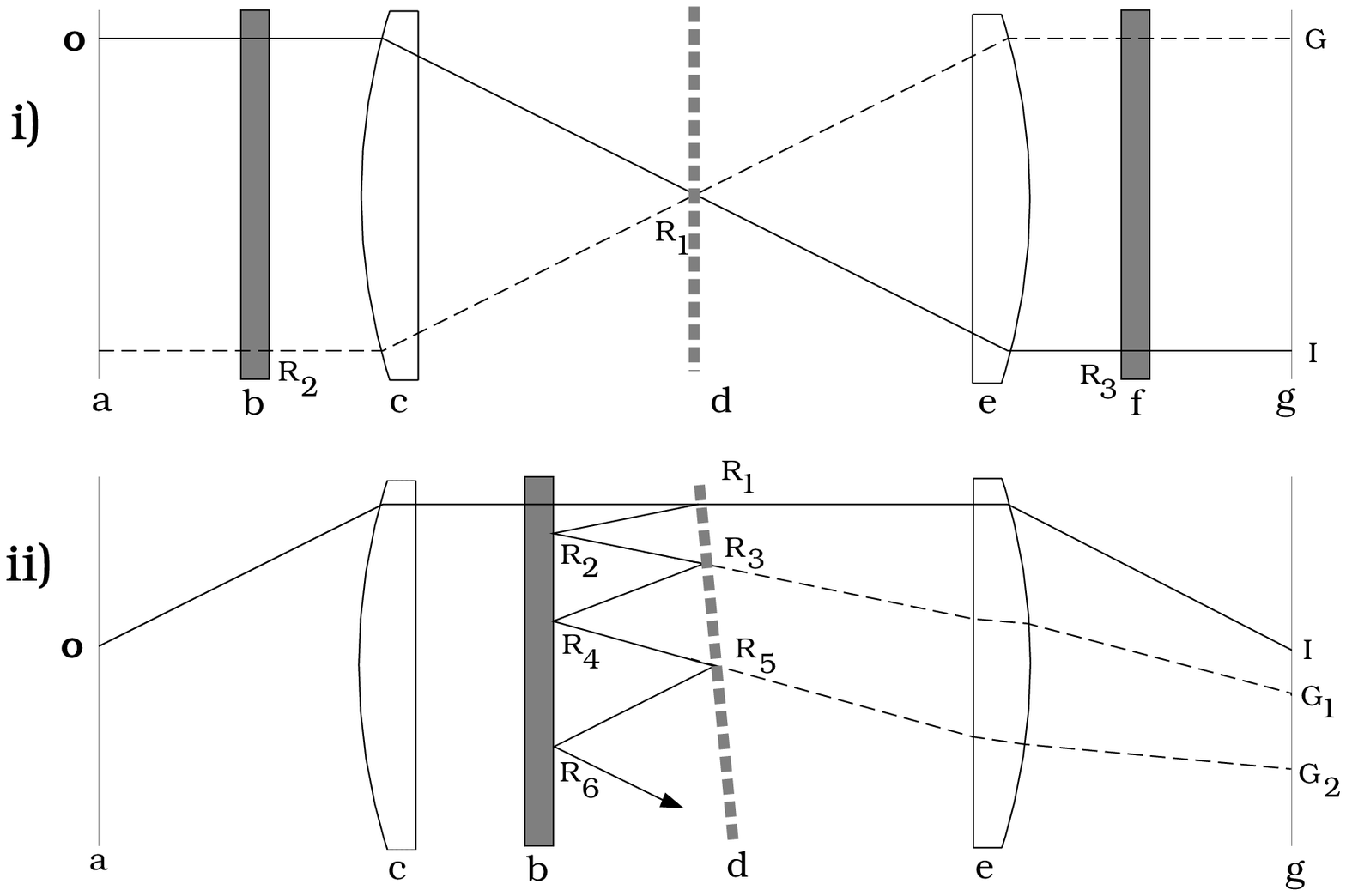}
}
\caption{
\singlesp
Ghost families arising from internal reflections within a Fabry-Perot spectrometer.
(i) Diametric ghosts. Rays from the object O form an inverted image I
and an out-of-focus image at R$_3$. The reflection at R$_1$ produces an out-of-focus
image at R$_2$. The images at R$_2$ and R$_3$ appear as a ghost image G at the detector.
(ii) Exponential ghosts. The images at R$_2$ and R$_4$ appear as ghost images G$_1$ and G$_2$
respectively.
}
\end{figure}

It is possible to radically alter the resolving power and free spectral range by scanning 
etalons in series simultaneously [46]. In principle,
all possible combinations of a mixture of $3-5$ high and low finesse etalons can
be used to mimic the effect of a tunable filter. While it is still necessary to use a
broadband filter to block unwanted orders, only a handful are needed to cover the
full optical spectrum. However, to
use a Fabry-Perot as a tunable filter without the necessary phase correction requires
that we restrict the observations to the Jacquinot central spot. This is defined as
the field about the optical axis within which the peak wavelength variation with field
angle does not exceed $\sqrt{2}$ of the etalon band-pass [21].
A good discussion of Fabry-Perot based tunable filters is given by [39]. The basic principle
is to use the etalon in low order so that widely different wavelength regions are accessible
by using the adjacent orders. At optical and near-infrared wavelengths, Eq.\ (11) indicates
that to reach the lowest orders requires gap spacings of only a few microns. A major obstacle,
however, is that special techniques are now needed to deposit non-laminar dielectric 
coatings [38].  There have been recent advances in maintaining $\lambda/200$
parallelism over plate separations of 30 mm down to 1.8 \micron\ [39], allowing
for free spectral ranges of 0.05$-$50 nm at optical wavelengths. Presently,
techniques are under investigation for purging incompressible dust grains
from the etalon air gap. Thus, in principle, it should be possible to operate several
etalons at low order, both separately and in tandem, to simulate a variable band-pass 
tunable filter over the visible or near-infrared spectrum. However,
the design, manufacture and stability of extended bandwidth coatings remain important obstacles
for broadband tunable filters. 

Even a minimal Fabry-Perot arrangement can have eight or more optically flat surfaces.
At some level, all of these surfaces interact separately to generate spurious reflections.
The periodic behavior of the etalon requires that we use a narrowband filter somewhere in
the optical path. Typically, the narrowband filter is placed in the converging beam before
the collimator or after the camera lens (Fig. 5(i)). The filter introduces ghost reflections 
within the Fabry-Perot optics (Fig. 5(ii)). The pattern of ghosts imaged at the detector is
different in both arrangements, as illustrated in Fig.\ 5. The dominant reflections are mostly
deflected out of the beam by tilting the etalon through a small angle with respect to
the optical axis. A more difficult problem arises from the optical blanks which form the basis of
the etalon.  These can act as internally reflecting cavities (cf. Fig. 4(i)) that generate a high 
order Airy pattern at the detector [18,43]. Traditionally, the outer surfaces have been 
wedge-shaped to deflect this spurious signal out of the beam (Fig. 4(iii)). Even curved lens 
surfaces occasionally produce `halation' around point source images which may require experimenting
with both bi-convex and plano-convex lenses when designing a focal reducer.

\hh{    18.5.3.}{Birefringent Filter.}
The underlying principle of birefringent filters is that light originating in a single polarization 
state can be made to interfere with itself [47]. The Michelson interferometer ($\S$18.6) achieves 
interference by splitting the input beam and sending the rays along different path lengths before 
recombining them. By analogy, an optically anisotropic, birefringent medium can be used to produce 
a relative delay between ordinary and extraordinary rays aligned along the fast and slow axes of 
the crystal.  (A birefringent medium has two different refractive indices, depending on the
plane of light propagation through the medium.) Title and collaborators have discussed at length the 
relative merits of different types of birefringent filters [see references in 47]. 
The filters are characterised by a series of perfect polarizers 
(Lyot filter [48,49]), partial polarizers, or only an entrance and an exit polarizer (Solc filter [50]). 
The highly anisotropic off-axis behaviour of uniaxial crystals give birefringent filters a major
advantage. Their solid acceptance angle is one to two orders of magnitude larger
than is possible with interference filters ($\S$18.5.1) although this is partly offset by half the 
light being lost at the entrance polarizer. To our knowledge, there has been no attempt to construct
a birefringent filter with a polarizing beam-splitter, rather than an entrance polarizer, in order to 
recover the lost light.

The Lyot filter is conceptually the easiest to understand. The entrance polarizer is oriented
45$^{\circ}$ to the fast and slow axes so that the linearly polarized, ordinary and extraordinary rays 
have equal intensity. The time delay through a crystal of thickness $d$ of one ray with respect to the 
other is simply $d\ \Delta\mu / c$
where $\Delta \mu$ is the difference in refractive index between the fast and slow axes. 
The combined beam emerging from the exit polarizer shows intensity variations described
by $I^2\cos(2\pi\ d\ \Delta\mu/\lambda)$ where $I$ is the wave amplitude. As originally illustrated by
Lyot [49], we can isolate an arbitrarily narrow spectral band-pass by placing a number of
birefringent crystals in sequence where each element is half the thickness of 
the preceding crystal. This also requires the use of a polarizer between each crystal 
so that the exit polarizer for any element serves as the entrance polarizer for the next.
The resolution of the instrument is dictated by the thickness of the thinnest element.

The instrumental profile for a Lyot filter with $s$ elements is
\begin{equation}
{\cal L} = {{1}\over{4^s}} {{\sin^2 \left(2^s\pi\ d\ \Delta \mu /\lambda\right)}\over{\sin^2 \left(\pi\ d\ \Delta \mu/\lambda\right)}}
\end{equation}
for which $d$ is now the thickness of the thinnest crystal element. 
By analogy with the Fabry-Perot ($\S$18.5.2), if $\lambda_0$ is the wavelength of the peak transmission,
the filter bandwidth ($= 0.88 \lambda^2_0 /(2^s d\ \Delta\mu)$), the free spectral range 
($= \lambda^2_0 /(d\ \Delta\mu)$) and the effective finesse ($= 1.13\ 2^s$) are easily derived.

It should be noted that $\lambda_0$ can be tuned over a wide spectral range by rotating the crystal 
elements. But to retain the transmissions in phase requires that each crystal element be rotated about
the optical axis by half the angle of the preceding thicker crystal. The NASA Goddard Space Flight Center 
have recently produced a Lyot filter utilising eight quartz retarders with a 13 cm entrance window. The
retarders, each of which are sandwiched with half-wave and quarter-wave plates in 
addition to the polarizers, are rotated independently with stepping motors under computer control. 
They achieve a band-pass of 4$-$8\AA\ tuneable over the optical wavelength range (3500$-$7000\AA).

\hh{    18.5.4.}{Acousto-optic Filter.}
In 1969, Harris and Wallace introduced a new type of electronically tunable filter that makes
use of collinear acousto-optic diffraction in an optically anisotropic medium [51]. Acousto-optic
tunable filters (AOTF) are formed by bonding piezo-electric transducers such as lithium niobate to 
an anisotropic birefringent medium. 
The medium has traditionally been a crystal, but polymers have been
developed recently with variable and controllable birefringence.  When the transducers are excited at
frequencies in the range 10--250 MHz, the ultrasonic waves vibrate the crystal lattice to form a moving
phase pattern that acts as a diffraction grating.  A related approach is to use liquid crystals made
of nematic (anisotropic) molecules. When a transverse electric field is applied, the molecules align
parallel to the field because of their positive dielectric anisotropy and form a uniaxial birefringent
layer.  In acousto-optic filters, the incident light Bragg-scatters off the moving pattern from one
polarization state into its orthogonal state.  The birefringent interaction permits a large
angular aperture which is unattainable with isotropic Bragg diffraction.  If the crystal is
thick enough and the driving power high enough (several watts for each cm$^2$ of aperture),
only a limited band of optical frequencies in the incident light is cumulatively diffracted for a
given acoustic frequency.  The wavevector of the diffracted output beam is the vector sum of the
input wavevector and the acoustical wavevector; out-of-band frequencies remain undeviated.
Crystals such as TeO$_2$, quartz and MgF$_2$ are highly transmitting and have an efficient
acousto-optic response in different wavelength regimes from 200 nm to 5 \micron.
For currently attained crystal homogeneity and thickness, \RES\ can reach 10$^4$ and
out-of-band light is rejected down to a level of about 10$^{-5}$.

\begin{figure}[t]
\label{AOTF}
\centering
\hbox{
\epsfxsize=3.8in
\epsfbox[-50 113 400 705]{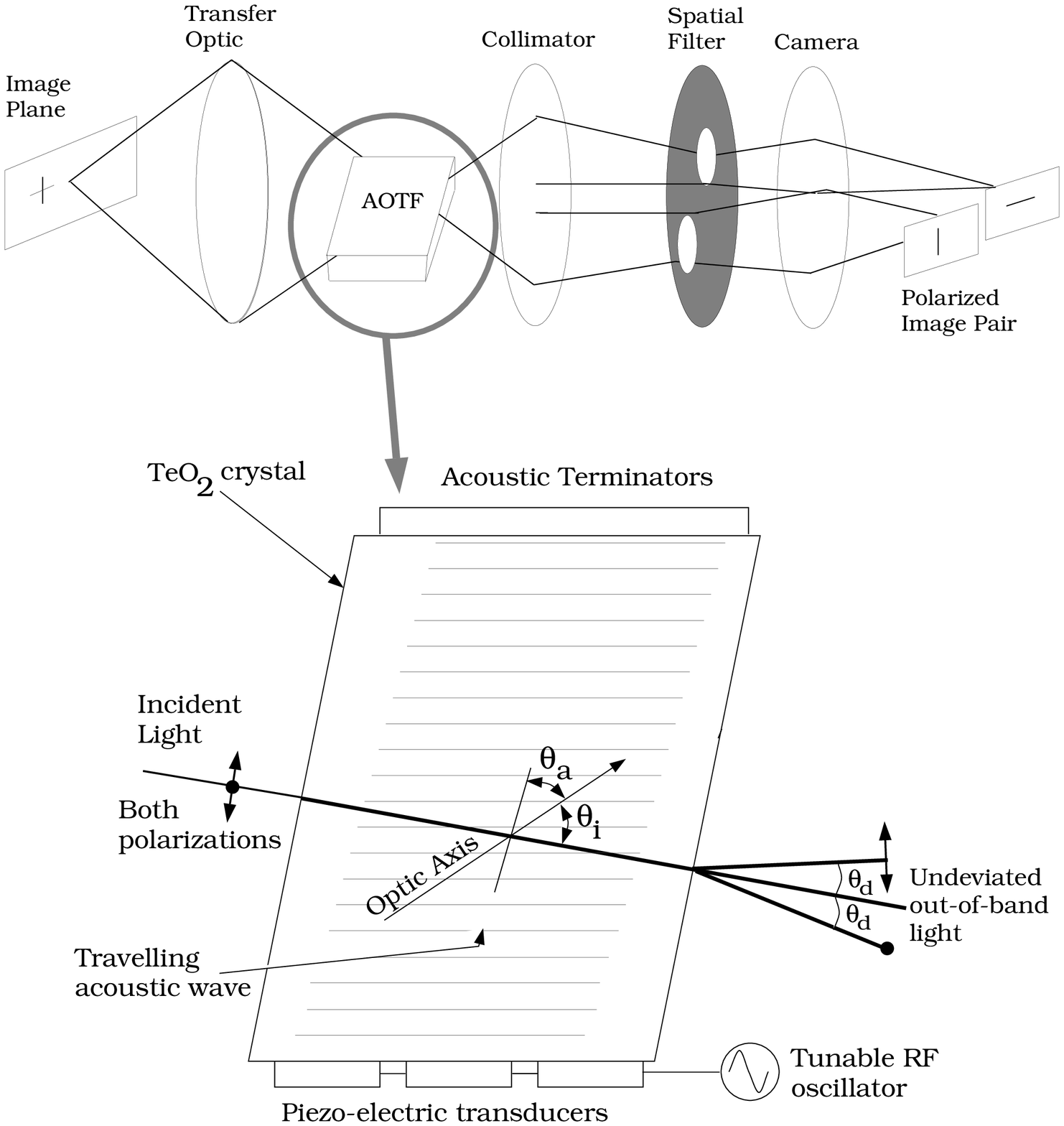}
}
\caption{
\singlesp
A crystalline, non-collinear acousto-optic tunable filter. $\theta_i$ is the
angle of incidence relative to the optic axis, $\theta_a$ the angle between the
acoustic wave and optic axis, and $\theta_d(\lambda)$ the diffraction angle.
The beams separate when they exit the filter.
}
\end{figure}

In the more useful
non-collinear filters [52,53], the acoustic and optical wave vectors differ in such a way that the
phase differences introduced by variations in the angle of incidence can be approximately compensated 
by the different refractive indices of the ordinary and extraordinary rays
$n_o$ and $n_e$, respectively.  For an extraordinary polarized incident beam,
$n_i = ((\cos\theta_i/n_o)^2+(\sin\theta_i/n_e)^2)^{-1/2}$.
If the incident
angle is $\theta_i$, the vacuum central wavelength is $\lambda_0$, and the 
acoustic wavelength is
$\Lambda$, the diffracted angle $\theta_d$ depends only weakly on $\lambda$
and is given by solving
\begin{equation}
\left(\frac{\lambda_0}{\Lambda}\right)^2 = 
{n_o}^2+{n_i}^2-2n_o n_i \cos [\theta_i-\theta_d(\lambda)]
\end{equation}
If $\tan \theta_d = (n_o/n_e)^2 \tan \theta_i$ (the non-critical phase matching
configuration), and the interaction length is short enough,
the acceptance angle for diffraction-limited imaging
can increase to almost 28\deg\ (i.e. an f/2 beam)
to provide throughput comparable to that of a Fabry-Perot.
However, \RES\
is then limited by the small attainable density of the acoustic driving power.
The diffracted ordinary and extraordinary rays emerge
displaced to either side of the undiffracted
out-of-band light by angles of 1\deg--10\deg\ (Fig.\ 6).
The undeviated out-of-band light is removed with an aperture mask in the collimated beam. 
It is particularly difficult at infrared wavelengths to keep this light from scattering back
into the optical path, thereby increasing the background.  While the corresponding
f/numbers of 50 to 5 are too slow for imaging spectroscopy of low surface-brightness sources,
for brighter objects these filters simultaneously deliver the two
orthogonally polarized images with more than $65\%$ efficiency and so are reliable and
compact imaging spectro{\it polarimeters} [54].
In beams as fast as f/2, the incident light can be sent through a linear
polarizer in front of the filter, and a crossed polarizer removes one of the
deviated beams after passage through the crystal for a net throughput in excess of $30\%$.

The instrumental profile of an acousto-optic filter of length $L$ that operates at
a central wavelength $\lambda_0$ in a collimated beam is like that of a grating,
\begin{equation}
T = T_0\ {\sf sinc}^2 \frac{\Delta k\ L}{2 \pi},
\end{equation}
with FWHM
\begin{equation}
\Delta\lambda = \frac{0.9\pi\lambda_0^2}{L d \sin^2\theta_i} = \frac{2 \pi n}{\Delta k},
\end{equation}
where $L$ is the interaction length, $n$ is one
index of refraction, $\Delta k$ is the mismatch between optical and acoustic wavevectors,
and $\theta_i$ is the incident angle relative to the crystal optical axis.
The dispersive term, $d(\lambda)$, becomes very large near a band edge.
The band-pass can be altered rapidly across a large wavelength range merely by
tuning the power and frequency of the acoustic wave, to form a composite band-pass
shape with widely separated, broad multiple peaks.
Such spectral multiplexing across several selected pass bands is a unique capability of
acousto-optic filters.

These filters can be used alone as moderate-resolution imaging spectrometers, even in the
ultraviolet and infrared where it is extremely difficult to 
produce interference filters with band widths narrower than $\lambda$/100 which also have
high transmission and good off-band rejection.  Alternatively, they can be used
in a collimated beam as the order-sorting filter of a Fabry-Perot etalon.
The spatial resolution can be as good as 15 \micron, well matched to the size of CCD
detector elements.  Crystal performance does not appear to deteriorate with age, unlike 
multi-layer coatings used with interference filters and Fabry-Perot etalons.

Current disadvantages include their expense, long fabrication time, and small size
($<$25 mm square for crystals that are uniform enough for good imaging) relative to interference
filters, restrictions that are likely to be lifted as the commercial market develops.  Another
concern that is particularly acute in the infrared is the power dissipated and heat generated during
their operation.  Nonetheless, an acousto-optic filter today is often much cheaper than a comparably
performing Fabry-Perot and control electronics, and also avoids the
complications of selecting among the multiple spectral
orders that characterize etalons.

\subsection
{Two Beam Interferometers}

All spectroscopic techniques rely ultimately on the interference of beams that traverse different
optical paths to form a signal. The prism uses essentially an infinite number of beams 
whereas the grating uses a finite number of beams set by the number of grooves. The
Fabry-Perot uses a smaller number of beams set by the instrumental finesse.
Bell [19] notes that as the number of beams decreases, the throughput (and therefore efficiency) of
the spectrograph increases.  Because at least two beams are required for interference, Bell concludes
that two-beam interferometers are the ultimate in spectrometers. The two most commonly used Fourier
Transform devices divide either the wavefront (lamellar grating interferometer) or the wave amplitude 
(Michelson interferometer). The efficiency of the latter is $\la$50\% whereas the former
technique can approach 100\%. Lamellar gratings are discussed in [13].

\begin{figure}[t]
\label{FTS}
\centering
\hbox{
\epsfxsize=3.5in
\epsfbox[0 225 450 760]{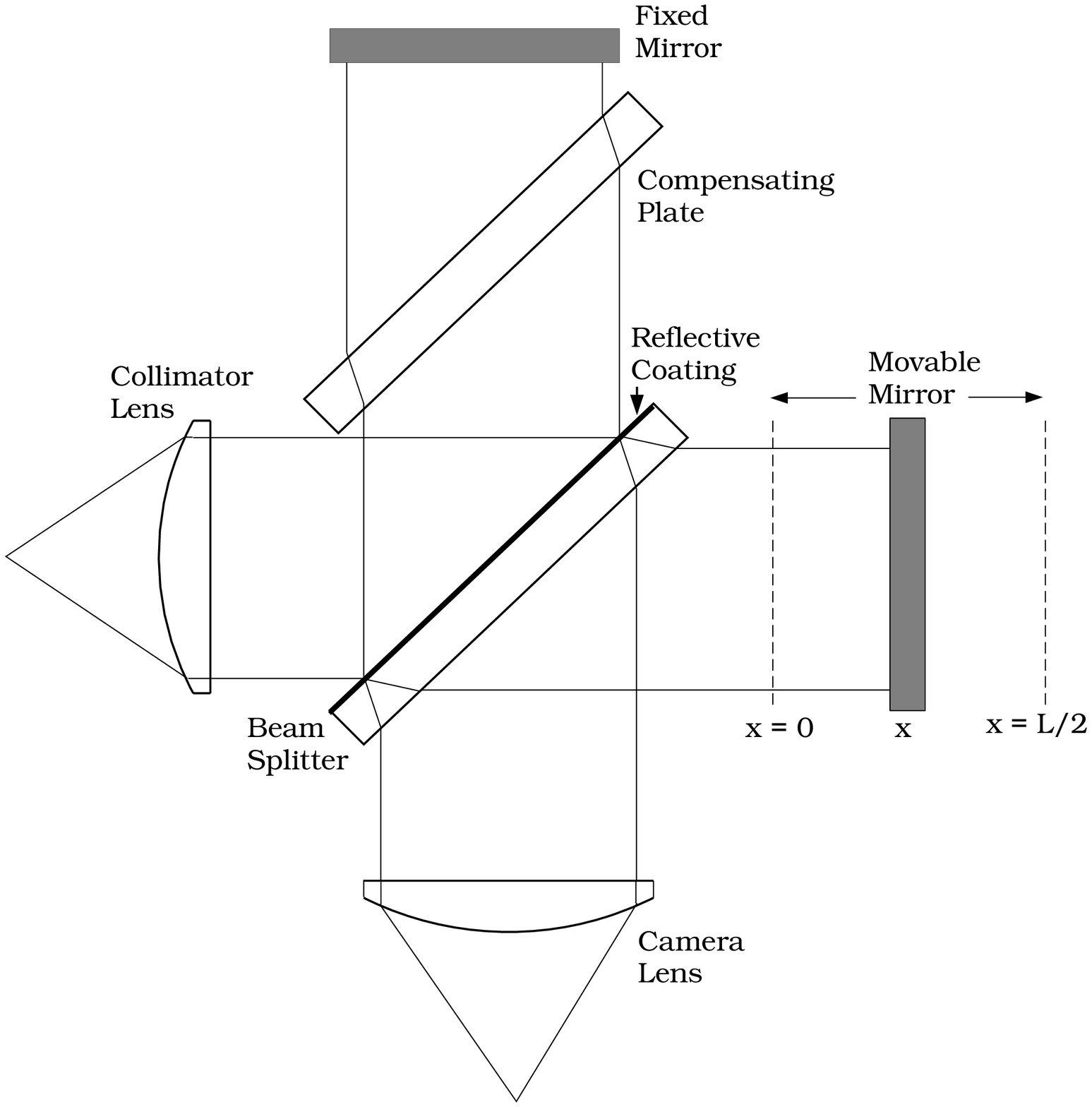}
}
\caption{
\singlesp
Two-beam Michelson interferometer (Fourier Transform spectrometer).
}
\end{figure}

A simple two-beam Michelson interferometer is shown in Fig.\ 7 and forms the basis of
the Fourier Transform spectrometer. The collimated beam is split into two beams at the
front surface of the beam-splitter. These beams then undergo different path lengths by
reflections off separate mirrors before being imaged by the camera lens at the detector.
The device shown in Fig.\ 7 uses only 50\% of the available light. It is possible to
recover this light but the layout is involved [55,56]. For all systems,
the output signal is a function of path difference between the mirrors. At zero path
difference (or arm displacement), the waves for all frequencies interact coherently. 
As the movable mirror is scanned, each input wavelength generates a series of transmission 
maxima.  Commercially available devices usually allow the mirror to be scanned continuously 
at constant speed, or to be stepped at equal increments. At a sufficiently large arm 
displacement, the beams lose their mutual coherence.

The spectrometer is scanned from zero path length ($x=y=0$) to a maximum path length $y=L$
set by twice the maximum mirror spacing ($x=L/2$). The superposition of two coherent beams with
amplitude $b_1$ and $b_2$ in complex notation is $b_1 + b_2 e^{i 2\pi\nu y}$ where $y$ is the 
total path difference and $\nu$ is the wavenumber. If the light rays have the same intensity, the 
combined intensity is $2 b^2 (1+\cos 2\pi\nu y)$, or equivalently, $4 b^2 \cos^2 \pi\nu y$, where 
$b = b_1 = b_2$. The combined beams generate a series of intensity {\it fringes} at the detector.
If it was possible to scan over an infinite mirror spacing at infinitesimally small spacings of the 
mirror, the superposition would be represented by an ideal Fourier Transform pair, such that
\begin{eqnarray}
b(y)   &=& \int^{\infty}_{-\infty} B(\nu) (1+\cos 2\pi\nu y)\ d\nu \\
B(\nu) &=& \int^{\infty}_{-\infty} b(y) (1+\cos 2\pi\nu y)\ dy
\end{eqnarray}
where $b(y)$ is the output signal as a function of pathlength $y$ and $B(\nu)$ is the spectrum we 
wish to determine. $B(\nu)$ and $b(y)$ are both undefined for $\nu < 0$ and $y<0$: we include the 
negative limits for convenience. Notice that 
\begin{eqnarray}
b(y)-{{1}\over{2}}b(0) &=& \int^{\infty}_{-\infty} B(\nu) \cos 2\pi\nu y\ d\nu \\
B(\nu) &=& \int^{\infty}_{-\infty} [b(y)-{{1}\over{2}}b(0)] \cos 2\pi\nu y\ dy
\end{eqnarray}
The quantity $b(y)-{{1}\over{2}}b(0)$ is usually referred to as the {\it interferogram} although
this term is sometimes used for $b(y)$. The spectrum $B(\nu)$ is normally computed using widely
available Fast Fourier Transform methods.

It is clear that the output signal is sinusoidal about some mean continuum value.
If the wavefronts have different intensities, the depth of the modulation decreases
and the mean continuum level increases. This is undesirable because the background
continuum constitutes a source of noise. The contrast of the fringe amplitude with respect to the
background is known as the {\it fringe visibility}. It is particularly important to ensure
that all rays undergo the same optical path, otherwise the output signal is asymmetric with
respect to zero path length. In Fig.\ 7, the beam that reflects off the movable
mirror passes through the beam-splitter three times. The compensating plate ensures that
the beam that reflects off the fixed mirror undergoes the same optical path. 

Efficient beam-splitters are crucial to the operation of an FTS device. These often comprise
dielectric sheets, wire grids or multi-layer dielectric coatings on substrates (\S~18.5.1),
depending on the wavelength of operation. Particular care must be taken over the internally
reflecting rays. Typically, the primary transmitted ray dominates and the secondary transmitted
ray can be neglected. However, the primary and secondary {\it reflected}
rays (cf. Fig. 4(i)) are
comparable in intensity. To maximize peak efficiency this requires a beam-splitter which maintains constructive interference 
between the rays over as much of the wavelength range as possible.

In practice the ideal Fourier Transform pair is not realized. 
The finite maximum baseline $y=L$ 
set by the maximum mirror displacement ultimately limits the instrumental resolution although, 
under certain circumstances, the size of the source can impose a stronger constraint [19]. The
response of most
interferograms declines with higher wavenumbers. At some intermediate value of 
$y$, the signal-to-noise ratio may fall to unacceptably low levels, in which case the effective
resolution may be somewhat lower than the theoretical value. The FTS is scanned at discrete sampling 
intervals (mirror spacings) which causes the computed spectrum to replicate at wavenumber intervals 
that are inversely proportional to the sampling interval (\S~18.2.2). If the baseline $L/2$ is 
sufficiently large to resolve all details within the spectrum, the replicated spectra will not overlap. 
Thus, the integrals in Eqs. (18) and (19) should be replaced by summations over finite limits. 

The optical alignment of an FTS is particularly involved [19]. Traditionally, this is done
by using, say, a HeNe laser along the optical axis and aligning the beam-splitter with the movable
and fixed mirrors separately and then together. The mirrors must be aligned and sufficiently
flat so as to introduce no wavefront errors greater than $\lambda/8$ [13]. This is a major
opto-mechanical challenge over the large arm displacements ($\la$1 m) of a two-beam interferometer.
In some respects, the Fabry-Perot constraint of $\lambda/(2N)$ is easier to meet because the 
plates are optically contacted at a physical spacing of, say, 1 mm, and then scanned 
through a few orders about this spacing, a total mechanical distance of only a few microns.
There are many ways to introduce phase errors into the computed spectrum. This is particularly
so if the stepping does not start at the precise position for the zero path difference.

While attention has been given to the difficulties of two-beam interferometry, the disadvantages
are few when compared with other spectrometers. In particular, the instrumental profile can take 
any form after mathematical filtering. The interferogram of a single wavelength is given by 
$b(y) = B(\nu_0)\cos 2\pi \nu_0 y$. The expected spectrum is then
\begin{equation}
S(\nu) = B(\nu_0) \sum^{y=L}_{y=-L} \cos 2\pi\nu_0 y\ \cos 2\pi \nu y\ \delta y
\end{equation}
It is straightforward to show that this reduces to
\begin{equation}
S(\nu) = B(\nu_0)\ L\ ({\sf sinc}\ 2\pi(\nu_0-\nu)L + {\sf sinc}\ 2\pi(\nu_0+\nu)L)
\end{equation}
Because the second term is negligible ($<$1\%) at optical and infrared wavelengths, the basic 
instrumental profile is a {\sf sinc} function. This has the highest resolution possible for a 
baseline of $L$. 

At the expense of resolution, the convolution theorem [57] allows one to modify digitally the 
instrumental response, i.e., for an arbitrary digital filter $f(y)$, we can write
\begin{equation}
S(\nu) = {\sf FT}[f(y) b(y)] = F(\nu) \star B(\nu)
\end{equation}
where $\star$ denotes convolution, {\sf FT} is the Fourier Transform operator and 
$F(\nu) = {\sf FT}[f(y)]$. A common reason to modify the instrumental response is to reduce the 
side lobes of the {\sf sinc} function, a process known as {\it apodization} [57]. A list of
apodizing functions is given in [19]. In particular, if the interferogram is
convolved with a triangle 
function, $f(y)=1-\vert y\vert/L$, the instrumental profile now takes the form of a {\sf sinc}$^2$ function
which is the response of both the grating ($\S$18.4) and the acousto-optic filter ($\S$18.5.4).

\noindent
{\bf Acknowledgments.}

We wish to thank P.D. Atherton (Queensgate Instruments), S.C. Markham (MediMedia), 
J. O'Byrne (University of Sydney), P.L. Shopbell (Rice University and Caltech)
and Lady Anne Thorne (Blackett Laboratory) for critical readings of an early manuscript.

\end{document}